# Generalized Mathematical Formalism Governing Free-carrier Driven Kerr Frequency Comb in Optical Micro-cavities

Raktim Haldar[1,*], Arkadev Roy[1], Partha Mondal[2], Vishwatosh Mishra[2] and Shailendra K. Varshney[1]

1. Department of Electronics and Electrical Communication Engineering, Indian Institute of Technology, Kharagpur-721302, India
2. Department of Physics, Indian Institute of Technology, Kharagpur-721302, India

[*]raktim@ece.iitkgp.ernet.in

**Abstract:** Continuous-wave pumped optical microresonators have been vastly exploited to generate frequency comb (FC) utilizing the Kerr nonlinearity. Most of the nonlinear materials used to build photonic platforms exhibit nonlinear losses such as multi-photon absorption, free-carrier absorption (FCA), and free-carrier dispersion (FCD) which can strongly affect the nonlinear characteristics of the devices made out of these materials. In this work, we model the Kerr FC based on modified Lugiato-Lefever Equation (LLE) along with the rate equation and develop analytical formulations to make quick estimations of the steady-state, modulation instability (MI) gain, bandwidth and the dynamics of Kerr Frequency-Comb (FC) in presence of nonlinear losses. Our analytical model is valid over a broad wavelength range of interest as it includes the effects of all nonlinear losses. Higher order (>3) characteristic polynomial of intra-cavity power describing the steady-state homogeneous solution of the modified LLE are discussed in detail. We derive the generalized analytical expressions for the threshold of normalized pump detuning to initiate the optical bistability which is a necessary condition for the FC generation. Free-carrier dispersion-led nonlinear cavity detuning is observed through the reverse Kerr-tilt of the resonant-peaks. We further deduce the expressions for the threshold pump intensity and the range of possible cavity detuning for the initiation of the MI when all the nonlinear losses are present. To corroborate our analytical findings, LLE along with the rate equations are solved numerically through split-step Fourier method. Our theoretical study can explain several experimental results which are previously reported and thereby is able to provide a better understanding of the comb dynamics.

## I. Introduction

Optical frequency-comb (FC) is a set of equidistant and coherent frequency lines in the ultraviolet, visible, and infrared regions [1], which can be used in precision measurement [2], microwave signal synthesis [3, 4], WDM-communication [5, 6], sensing [7], spectroscopy and molecular fingerprinting [2, 8], astronomy [9, 10], entangled photon pair generations [11] or as an optical ruler [1, 5]. Mode-locked femtosecond lasers and fiber lasers have extensively been used for the generation of optical FC [12, 13] until recently the parametric frequency conversion using continuous-wave (CW) optical pump in optical cavities revolutionizes the technology for comb generation [14, 15]. Schematic of optical cavities such as whispering gallery mode (WGM) micro-sphere, micro-disks and microring resonator are shown in Fig. 1. Eventually, the device foot-print has been reduced to a few hundred micrometers while a repetition rate as high as >10 GHz can easily be achieved [14, 15]. Varieties of materials have been used along with different novel fabrication techniques to design ultra-high quality factor micro-resonators [16-18] for low-threshold, stable FC generation, however; the search for the most suitable material is still on. FC generation in crystalline fluorides and $MgF_2$ [19-23], Hydex glass [24], diamond [25], quartz [26], aluminum-nitride [27, 28], lithium niobate [29] AlGaAs [30], silica [14], and silicon-nitride [31] have already been demonstrated. Apart from this, FC is also demonstrated in organically modified silica micro-cavities by Shen *et al.* in 2017 [32] with a very low threshold power.

Silicon-based platforms are often preferred due to several advantages such as tight optical confinement, high Kerr-coefficient, transparency over a broad wavelength range (from telecom to mid-IR), low-cost and most importantly its compatibility with the existing microelectronics industry [33]. High refractive index contrast between silicon and other cladding materials (air, silica) results in strong optical confinement in silicon waveguides which allows sharp waveguide bends that help further to reduce the device footprint [34]. Tight confinement also enhances the effective nonlinearity which facilitates the realization of different nonlinear phenomena with a very low input power [35-37]. It has been shown that efficient dispersion engineering in slot waveguides (made of silica, silicon, or silicon-nitride) has potential to achieve broadband frequency-combs [38]. On the other hand, silicon exhibits strong two-photon absorption (2PA/TPA) below < 2.2 μm wavelength [39-41] which can be exploited to realize all-optical logic operations [42] as well as all-optical signal processing [43]. However, high 2PA in silicon effectively reduces the figure of merit of silicon in nonlinear applications [36] thereby limiting us to use silicon in most of the nonlinear and quantum applications in the telecommunication and near-IR (NIR) wavelength despite having many advantages [44]. In short, mid-IR and far-IR wavelength ranges, silicon is also not free from three-photon absorption (3PA) and four-photon absorption (4PA) losses [45]. Therefore, due to its low nonlinear losses silicon-nitride ($Si_3N_4$) is often a preferred choice over silicon in nonlinear applications [46] although with an expense of Kerr coefficient which is about ten times lower than that of the silicon. Recently, octave-spanning frequency combs have been demonstrated both experimentally and numerically in $Si_3N_4$ [47] and Si microring resonators (MRRs) [48, 49]. Thus, a more realistic theoretical study on comb dynamics for Si-MRRs, applicable for a broad wavelength range of operation becomes indispensable where all the nonlinear losses and higher-order dispersion terms are being considered.

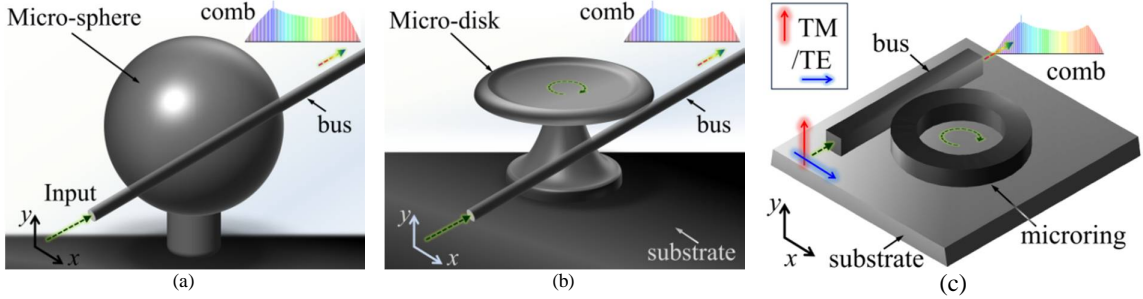

Fig. 1. Schematic of various optical micro-cavities which can be used as a platform for FC generation: (a) WGM microsphere, (b) WGM micro-disk, (c) Microring Resonators.

Nonlinear losses include multi-photon absorptions (2PA, 3PA, 4PA etc.), free-carrier absorption (FCA) and free-carrier dispersion (FCD). Generalized mean-field Lugiato-Lefever equation (LLE) can be used to model the Kerr FC in a high-$Q$, high-finesse optical cavity where the dispersion, nonlinear phase accumulation over a round-trip and pump detuning is low [50]. It is known that temporal cavity soliton (CS) generates Cherenkov radiation (dispersive-wave) in presence of higher-order dispersions which leads to octave-spanning FC [46]. Recently, numerical studies have shown that FCA-induced FCD causes nonlinear cavity detuning which in turn helps to generate optical FC even in absence of linear-detuning of the CW-pump [51]. 2PA in telecom/NIR and 3PA with FCA, FCD in short and mid-IR wavelength range generally inhibit the parametric oscillation in silicon-waveguide [45]. To combat free-carrier induced nonlinear losses, external reverse-bias has been employed across the PIN junction [48] which has been fabricated along the cross-section of the Si-waveguide. The external bias minimizes the FCA-FCD effects by sweeping the generated free-carriers. This method facilitates the broadband (2.1μm-3.5μm [48] and 2.4μm-4.3μm [52]) FC within the mid-IR wavelength range. Breather solitons have been demonstrated both in $Si_3N_4$ and Si-waveguides and the effect of 3PA along with the corresponding FCA-FCD is taken into consideration in the theoretical models [52, 53]. However, a detailed theoretical study that includes the effects of all nonlinear losses and free-carriers on the generation of Kerr-comb is still scarce.

In this work, for the first time to the best of our knowledge, we find the steady state homogeneous solutions of free-carrier driven generalized mean-field LLE [54] and report the reverse Kerr-tilt as a consequence of FCA-FCD induced nonlinear cavity detuning [55]. The characteristics polynomials for the steady-state homogeneous solutions of the LLE possessing all the nonlinear losses are derived having an order greater than three instead of the well-known cubic polynomial [56]. In subsequent sections, we discuss the threshold detuning to initiate the bistability in presence of multi-photon absorption and free-carrier effects which is a necessary condition to obtain FC. Finally, we generalize the existing formulations [56-58] to study the modulation instability (MI) gain and bandwidth in presence of higher-order dispersion terms, 2PA, 3PA, 4PA, FCA, and FCD. Most of the parameters used in simulations are taken from [45]. To validate the analytical model we solve the modified LLE along with the coupled rate equation numerically through split-step Fourier method. Our theoretical study can explain several experimental results [45, 48, 52] and thereby provides an in-depth understanding of the FC dynamics in the most practical scenarios.

**II. Theoretical Model**

Micro-resonator based Kerr FC can be modeled by mean-field LLE. Solving the LLE is computationally less intensive than other methods [50, 59] while the numerical results obtained from LLE match with the experiments reasonably well even for octave-spanning FC-s.

*A. Normalization of LLE*

The generalized mean-field Lugiato-Lefever equation including multi-photon absorption, FCA, and FCD to model the Kerr-comb along with the rate equation can be given as [45]:

$$t_R \frac{\partial E(t,T)}{\partial t} = -(\alpha + i\delta_0)E(t,T) + \left[iL\sum_{k\geq 2}\frac{\beta_k}{k!}\left(i\frac{\partial}{\partial T}\right)^k + \left(1 + \frac{i}{\omega_0}\frac{\partial}{\partial T}\right)\left(i\gamma L|E(t,T)|^2 - \frac{\beta_{2PA}L}{2A_{eff}}|E(t,T)|^2 - \frac{\beta_{3PA}L}{3A_{eff}^2}|E(t,T)|^4 - \frac{\beta_{4PA}L}{4A_{eff}^3}|E(t,T)|^6\right)\right]E(t,T)$$
$$- \frac{\sigma L}{2}(1+i\mu_c)N_c(t,T)E(t,T) + \sqrt{\theta}E_{in} \quad (1)$$

$$\frac{\partial N_c(t,T)}{\partial T} = \frac{\beta_{2PA}}{2\hbar\omega_0}\frac{|E(t,T)|^4}{A_{eff}^2} + \frac{\beta_{3PA}}{3\hbar\omega_0}\frac{|E(t,T)|^6}{A_{eff}^3} + \frac{\beta_{4PA}}{4\hbar\omega_0}\frac{|E(t,T)|^8}{A_{eff}^4} - \frac{N_c(t,T)}{\tau_{eff}} \quad (2)$$

Where $\tau$, $T$ and $t_R$ represent the fast-time, slow-time, and the round-trip time, respectively. $t_R=\text{FSR}^{-1}$ if FSR is the free-spectral range of the cavity. $E(t,T)$ is the field amplitude whereas $\alpha$, $\delta_0$, $L$, $\beta_k$, $\omega_0$, $E_{in}$ denote the dimensionless total round-trip loss coefficient, external pump detuning, cavity perimeter length, $k$-th order dispersion coefficient, the angular frequency of the CW-pump and input pump amplitude, respectively. $\beta_{nPA}$ ($n$=2, 3, 4), $\sigma$, $\mu_c$, $N_c$, $A_{eff}$, $t_{eff}$, and $\hbar$ ($=h/2\pi$) are the multi-photon absorption coefficient, FCA, FCD, free-carrier density, effective mode area, carrier lifetime and reduced Planck's constant, respectively. At first, we neglect the self-steepening term and normalize [50] the LLE and the rate-equation including multi-photon absorption ($n$PA), FCA and FCD terms in convenient forms, as given by equations (3) and (4), respectively.

$$\frac{\partial u}{\partial \xi} = -(1+i\Delta)u - \frac{1}{2}(1+iK)\phi_c u - i\frac{s}{2}\frac{\partial^2 u}{\partial \tau^2} + \sum_{k\geq 3}\frac{d_k}{k!}\frac{\partial^k u}{\partial \tau^k} + i|u|^2 u - \frac{Q_2}{2}|u|^2 u - \frac{Q_3}{3}|u|^4 u - \frac{Q_4}{4}|u|^6 u + S \quad (3)$$

$$\frac{\partial \phi_c}{\partial \tau} = \theta_{c_2}|u|^4 + \theta_{c_3}|u|^6 + \theta_{c_4}|u|^8 - \frac{\phi_c}{\tau_c} \quad (4)$$

$u(\zeta,\tau)$, $\zeta$, $\Delta$, $K$, $\phi_c$, $s$, $d_k$, $Q_n$ ($n$=2, 3, 4), $S$, $\theta_{cn}$ ($n$=2, 3, 4) and $\tau_c$ are the normalized- intra-cavity field amplitude, slow-time, fast-time, pump detuning, FCD co-efficient, free-carrier density, second and higher order dispersion terms, multi-photon absorption coefficients, pump amplitude, FCA coefficients and normalized carrier lifetime, respectively where the normalization factors are given as,

$$u = \frac{E}{\psi} = \frac{E}{\sqrt{\frac{\alpha}{\gamma L}}}, \qquad \tau = \frac{T}{T_0}, \qquad T_0 = \sqrt{\frac{|\beta_2|L}{\alpha}}, \qquad \xi = \frac{t}{L_D}, \qquad L_D = \frac{t_R}{\alpha},$$

$$\Delta = \frac{\delta_0}{\alpha}, \qquad \phi_c = \frac{N_c}{\eta_c}, \qquad \eta_c = \frac{t_R}{\sigma LL_D}, \qquad \tau_c = \frac{\tau_{eff}}{T_0}, \qquad S = \sqrt{\frac{\theta\gamma L}{\alpha^3}}E_{in}$$

$$d_k = i^{k+1}\left(\frac{\alpha}{L}\right)^{\frac{k}{2}-1}\frac{\beta_k}{|\beta_2|^{\frac{k}{2}}}, \qquad Q_n = \frac{\beta_{nPA}LL_D}{A_{eff}^{(n-1)}t_R}\psi^{2(n-1)}, \qquad \theta_{c_n} = \frac{\beta_{nPA}T_0}{nh\upsilon_0\eta_c A_{eff}^n}\psi^{2n}, \qquad K = \frac{\sigma\eta_c LL_D\mu_c}{t_R},$$

The equations (3) and (4) are expressed in the more compact series form (equations (5) and (6)) where $n$ can run from 2 to 4. Equations (5) and (6) can further be simplified by assuming the dominant multi-photon absorption term only in the operating pump wavelength range. This assumption holds satisfactorily true when the nonlinear losses either inhibit the formation of FC or significantly reduce the span of the frequency comb. Nevertheless, equation (5) and (6) are more suitable if the FC is octave spanned. Thus the expressions given in (5) and (6) including only the dominant $n$PA, FCA and FCD and the general expression of rate equation are reduced in equation (7) and (8):

$$\frac{\partial u}{\partial \xi} = -(1+i\Delta)u - \frac{1}{2}(1+iK)\phi_c u - i\frac{s}{2}\frac{\partial^2 u}{\partial \tau^2} + \sum_{k\geq 3}\frac{d_k}{k!}\frac{\partial^k u}{\partial \tau^k} + i|u|^2 u - \sum_{n=2}^{4}\frac{Q_n}{n}|u|^{2(n-1)}u + S \quad (5)$$

$$\frac{\partial \phi_c}{\partial \tau} = \sum_{n=2}^{4}\theta_{c_n}|u|^{2n} - \frac{\phi_c}{\tau_c} \quad (6)$$

$$\frac{\partial u}{\partial \xi} = -(1+i\Delta)u - \frac{1}{2}(1+iK)\phi_c u - i\frac{s}{2}\frac{\partial^2 u}{\partial \tau^2} + \sum_{k\geq 3}\frac{d_k}{k!}\frac{\partial^k u}{\partial \tau^k} + i|u|^2 u - \frac{Q_n}{n}|u|^{2(n-1)}u + S \quad (7)$$

$$\frac{\partial \phi_c}{\partial \tau} = \theta_{c_n}|u|^{2n} - \frac{\phi_c}{\tau_c} \quad (8)$$

Where $n$=2 in case of 2PA/TPA is dominant, $n$=3 while 3PA is significant and $n$=4 while 4PA is the dominant nonlinear loss mechanism. Depending upon the band-gap energy of silicon (i.e. $E_g$~01.1 eV), only 2PA, 3PA or 4PA is significant at the operating wavelength $\lambda$<2.2μm (telecommunication and near IR wavelength), $\lambda$>2.2 μm to $\lambda$<3.3 μm (short wavelength IR) and $\lambda$>3.3μm (mid-IR), respectively, for silicon waveguides. However, the above-mentioned operating wavelength varies depending upon the doping concentration of the intrinsic material of the optical waveguide. It also varies material to material. Free-carriers are generated through multi-photon absorption which induces additional FCA losses to the system. FCA is associated with the FCD as free-carriers are also able to modify the refractive index of the propagating medium [45]. Particularly, if the intensity of the input incident pulse is very high, effects of FCA and FCD will be crucial. However, in suitable conditions such as low pulse energy or in presence of external bias that is able to sweep the free-carriers, the effects of FCA-FCD can be neglected [45]. Table 1 provides a detailed account of the values of different normalized parameters in various conditions (presence and absence of different nonlinear losses (NLL) and higher order dispersion (HOD)).

**Table 1. Values of different parameters in presence and absence of different nonlinear losses and higher order dispersions**

| Parameters | No NLL & no HOD | No NLL, HOD present | Only $n$PA | Only $n$PA and FCA | $n$PA, FCA and FCD without HOD | $n$PA, FCA and FCD with HOD |
|---|---|---|---|---|---|---|
| $\theta_{cn}$ | $=0$ | $=0$ | $\neq 0$ | $\neq 0$ | $\neq 0$ | $\neq 0$ |
| $\tau_c$ | $=0$ | $=0$ | $\to 0$ | $\neq 0$ | $\neq 0$ | $\neq 0$ |
| $C_n(=\theta_{cn}.\tau_c)$ | $=0$ | $=0$ | $\to 0$ | $\neq 0$ | $\neq 0$ | $\neq 0$ |
| $\phi_c$ | $=0$ | $=0$ | $\to 0$ | $\neq 0$ | $\neq 0$ | $\neq 0$ |
| $\frac{\partial \phi_c}{\partial \tau}$ | $=-\infty$ | $=-\infty$ | $\to -\infty$ | Finite | Finite | Finite |
| $Q_n$ | $=0$ | $=0$ | $\neq 0$ | $\neq 0$ | $\neq 0$ | $\neq 0$ |
| $K$ | $=0$ | $=0$ | $=0$ | $0$ | $\neq 0$ | $\neq 0$ |
| $d_k$ (k≥3) | $=0$ | $\neq 0$ | $=0$ | $0$ | $0$ | $\neq 0$ |

Realistic values of different parameters taken from [45] along with their calculated normalized values are provided in Table 2. In our simulations, the waveguide area and the effective area ($A_{eff}$) are equivalent [45, 60].

**III. Steady-state Stationary Homogeneous Solution**

We find the steady-state, stationary ($\partial u/\partial \xi = 0$) homogeneous ($\partial u/\partial \tau = 0$) solutions of equation (5) with the corresponding rate equation (6) in presence of all nonlinear losses. In steady-state the wave amplitude, $u$ follows the relationship with the pump amplitude, $S$ as given by (9).

$$u = \frac{S}{(1+i\Delta) + \frac{1}{2}(1+iK)\sum_{n=2}^{4} C_n |u|^{2n} - i|u|^2 + \sum_{n=2}^{4} \frac{Q_n}{n}|u|^{2(n-1)}} \quad (9)$$

If the intra-cavity power and input pump power are denoted by $Y (=|u|^2)$ and $X (=|S|^2)$, respectively, the steady-state, stationary, and homogeneous solution of LLE can be expressed by the characteristics polynomial of $Y$ having a degree of $(2n+1)$ and can be written as (10) where $n$ varies from 2 to 4 if 2PA, 3PA, and 4PA, all are present. Thus the characteristics polynomial satisfies a nonic polynomial of $Y$.

$$X = (1+\Delta^2)Y - 2\Delta Y^2 + Y^3 + 2\sum_{n=2}^{4} \frac{Q_n}{n}Y^n + (1+\Delta K)\sum_{n=2}^{4} C_n Y^{n+1} - K\sum_{n=2}^{4} C_n Y^{n+2} \\ + \sum_{n=2}^{4} \frac{Q_n^2}{n^2}Y^{2n-1} + \frac{1}{4}(1+K^2)\sum_{n=2}^{4} C_n^2 Y^{2n+1} + \sum_{n=2}^{4}\sum_{q=2}^{4} \frac{C_n Q_q}{q}Y^{n+q} \quad (10)$$

In case, only one out of 2PA, 3PA and 4PA is dominant along with the FCA and FCD, equation (10) can be simplified as:

$$X = (1+\Delta^2)Y - 2\Delta Y^2 + Y^3 + \frac{2}{n}Q_n Y^n + C_n(1+\Delta K)Y^{n+1} - KC_n Y^{n+2} \\ + \frac{Q_n^2}{n^2}Y^{2n-1} + \frac{C_n Q_n}{n}Y^{2n} + \frac{C_n^2}{4}(1+K^2)Y^{2n+1} \quad (11)$$

**Table 2. Simulation Parameters: Realistic and normalized values in different operating wavelengths**

| Physical Parameters | Different wavelengths | | | | | |
|---|---|---|---|---|---|---|
| Pump Wavelength ($\lambda_p$) | $\lambda_p$=1.56μm Dominant: TPA/2PA | | $\lambda_p$=2.4μm Dominant: 3PA | | $\lambda_p$=4.0μm Dominant: 4PA | |
| | Value | Normalized Value | Value | Normalized Value | Value | Normalized Value |
| Pump Power | P=1W | $X=\|S\|^2$ = 128.064 | P=1W | $X=\|S\|^2$ = 174.55 | P=0.35W | $X=\|S\|^2$ = 8.459 |
| Waveguide Area | $A_{eff}$=300×690 (nm$^2$) | | $A_{eff}$=500×1400 (nm$^2$) | | $A_{eff}$=500×2600 (nm$^2$) | |
| Ring radius | 100μm | | 100μm | | 100μm | |
| Linear Loss | $\alpha_{dB}$=1.4dB/cm | $\alpha$=0.0187 | $\alpha_{dB}$=0.7dB/cm | $\alpha$=9.21×10$^{-3}$ | $\alpha_{dB}$=0.7dB/cm | $\alpha$=9.21×10$^{-3}$ |
| Coupling Coefficient | k=0.0172 | | k=0.0083 | | k=0.0083 | |
| Kerr Coefficient | $n_2$=4×10$^{-14}$ (cm$^2$/W) | | $n_2$=7×10$^{-14}$ (cm$^2$/W) | | $n_2$=3×10$^{-14}$ (cm$^2$/W) | |
| Nonlinear Coefficient | $\gamma$=77.8296 (W$^{-1}$m$^{-1}$) | | $\gamma$=26.1799 (W$^{-1}$m$^{-1}$) | | $\gamma$=3.6249 (W$^{-1}$m$^{-1}$) | |
| 2PA/TPA coefficient | $\beta_{2PA}$=1.5 (cm/GW) | $Q_2$= 0.931056 | | | | |
| 3PA coefficient | | | $\beta_{3PA}$=0.02 (cm$^3$/GW$^2$) | $Q_3$ =8.734×10$^{-4}$ | | |
| 4PA coefficient | | | | | $\beta_{4PA}$=3×10$^{-5}$ (cm$^5$/GW$^3$) | $Q_4$ =6.165×10$^{-6}$ |
| FCA coefficient | $\sigma$=1.47×10$^{-21}$ (m$^2$) | | $\sigma$=3.48×10$^{-21}$ (m$^2$) | | $\sigma$=9.67×10$^{-21}$ (m$^2$) | |
| FCD coefficient | $\mu$=7.5 | $K=K_2$=7.5 | $\mu$=4.9 | $K=K_3$=4.9 | $\mu$=2.9 | $K=K_4$=2.9 |
| Carrier Life-time | $\tau_{eff}$=3(ns) | $\tau_{c2}$ =6.6314×10$^4$ | $\tau_{eff}$=5(ns) | $\tau_{c3}$ =7.752×10$^4$ | $\tau_{eff}$=5(ns) | $\tau_{c4}$ =7.752×10$^4$ |
| FSR | 226 (GHz) | | 226 (GHz) | | 226 (GHz) | |
| Round-trip time | $t_R$=4.4248 (ps) | | $t_R$=4.4248 (ps) | | $t_R$=4.4248 (ps) | |
| Slow-time normalization factor | $L_D$= 2.3629×10$^{-10}$ (s) | | $L_D$= 4.8027×10$^{-10}$ (s) | | $L_D$= 4.8027×10$^{-10}$ (s) | |
| Amplitude normalization factor | | $\psi$=0.619 | | $\psi$= 0.7484 | | $\psi$=2.011 |

Thus in presence of only multi-photon absorption (nPA, n=2, 3 or 4) whereas, FCA and FCD are negligible equation (11) can further be reduced to:

$$X = (1+\Delta^2)Y - 2\Delta Y^2 + Y^3 + \frac{2}{n}Q_n Y^n + \frac{Q_n^2}{n^2}Y^{2n-1} \tag{12}$$

Equation (13) is the steady-state stationary solution of LLE while multi-photon absorption along with FCA is considered while the effect of FCD can be neglected.

$$X = (1+\Delta^2)Y - 2\Delta Y^2 + Y^3 + \frac{2}{n}Q_n Y^n + C_n Y^{n+1} + \frac{Q_n^2}{n^2}Y^{2n-1} + \frac{C_n Q_n}{n}Y^{2n} + \frac{C_n^2}{4}Y^{2n+1} \tag{13}$$

Therefore, we show that the steady-state, stationary solutions become quantic, septic and a nonic polynomial of Y in presence of only 2PA, 3PA or 4PA along with the corresponding FCA-FCD and are given by equations (14), (15) and (16), respectively.

$$X = (1+\Delta^2)Y + (Q_2 - 2\Delta)Y^2 + \left\{1 + \frac{Q_2^2}{4} + C_2(1+\Delta K)\right\}Y^3 + C_2\left(\frac{Q_2}{2} - K\right)Y^4 + \frac{C_2^2}{4}(1+K^2)Y^5 \tag{14}$$

$$X = \left(1+\Delta^2\right)Y - 2\Delta Y^2 + \left(1+\frac{2}{3}Q_3\right)Y^3 + C_3(1+\Delta K)Y^4 + \left(\frac{Q_3^2}{9} - KC_3\right)Y^5 + \frac{C_3 Q_3}{3}Y^6 + \frac{C_3^2}{4}\left(1+K^2\right)Y^7 \quad (15)$$

$$X = \left(1+\Delta^2\right)Y - 2\Delta Y^2 + Y^3 + \frac{Q_4}{2}Y^4 + C_4(1+\Delta K)Y^5 - KC_4 Y^6 + \frac{Q_4^2}{16}Y^7 + \frac{C_4 Q_4}{4}Y^8 + \frac{C_4^2}{4}\left(1+K^2\right)Y^9 \quad (16)$$

Note that all these equations (10)-(16) are reduced to the well-known cubic polynomial of $Y$ if all the nonlinear losses are neglected [56]. $C_n$ is defined as the product of $\theta_{cn}$ and the $\tau_c$. Thus, the polynomial of degree $(2n+1>3)$ accounts for the steady-state homogeneous solutions of LLE for multi-photon ($n$PA) absorption (case A, B) in presence of FCA and FCD.

$$X = \left[\left(1+\Delta^2\right)Y - 2\Delta Y^2 + Y^3\right] \quad (17)$$

**IV. Optical Bistability and Kerr-tilt**

Different important features such as the threshold pump power and threshold pump detuning to initiate the turing pattern and eventually the stable comb formation can be retrieved from the bistability curve and the Kerr-tilt plot which are obtained from the steady-state homogeneous solution of LLE [56-58]. In addition, many dynamic behaviors of frequency comb can partially be understood through Kerr-tilt and the bistability curve as the cavity soliton solutions of LLE results from the coexistence of patterned and CW solutions [61].

*A. Threshold Condition to initiate Optical Bistability*

The minimum value of the normalized cavity detuning $\Delta$ to initiate the optical bistability in absence of all the nonlinear losses is $\sqrt{3}$ [57, 58]. Multi-photon absorption along with FCA-FCD changes the threshold value of the cavity detuning. It is shown that in presence of 2PA, FCA-FCD the steady state homogeneous solution of the LLE satisfies the quintic polynomial of $Y$. However, if the FCA and FCD are negligible, the polynomial reduces to a cubic polynomial having 2PA coefficient ($Q_2 \neq 0$, $C_2=0$, $K=0$) which can be given as:

$$X = \left(1+\Delta^2\right)Y + (Q_2 - 2\Delta)Y^2 + \left\{1 + \frac{Q_2^2}{4}\right\}Y^3 \quad (18)$$

In this case, it is possible to find out an analytical expression of the threshold value of the normalized cavity detuning that initiates the optical bistability. The cubic polynomial having the 2PA coefficient, $Q_2$ can be given by:

$$\Delta_\pm \geq \frac{8Q_2}{4-3Q_2^2} \pm \frac{\sqrt{3}}{4-3Q_2^2}\left(Q_2^2 + 4\right) \quad (19)$$

Negative values of $\Delta$ are neglected. We calculate the threshold value of $\Delta$ from equation (19) and plot in Fig. 2 (a). It is observed that 2PA increases the threshold for optical bistability. We also find the values of the positions of the saddle nodes $X_\pm$ at which the bistability occurs and plot it in Fig. 2 (b). The values of $X_\pm$ are obtained using equation (20).

$$X_\pm = \pm\frac{4}{27}\left(\frac{1}{4+Q_2^2}\right)^2 \left(\pm 4\Delta \mp 2Q_2 + \sqrt{4-3Q_2^2}\sqrt{\Theta}\right) \quad (20)$$

$$\left(12 + Q_2^2 + \Delta^2\left(4+3Q_2^2\right) \pm Q_2\sqrt{4-3Q_2^2}\sqrt{\Theta} + 2\Delta\left(4Q_2 \mp \sqrt{4-3Q_2^2}\sqrt{\Theta}\right)\right)$$

$$\Theta = \Delta^2 - \frac{16Q_2}{4-3Q_2^2}\Delta + \frac{Q_2^2 - 12}{4-3Q_2^2} \quad (21)$$

In absence of 2PA equation (20) reduces to:

$$X_\pm = \pm\left(\frac{2}{27}\right)\left(\pm 2\Delta + \sqrt{\Delta^2 - 3}\right)\left(3 + \Delta^2 \mp \Delta\sqrt{\Delta^2 - 3}\right) \quad (22)$$

The intra-cavity steady-state power, $Y_\pm$ in presence of 2PA can be given by,

$$Y_\pm = -\frac{1}{3}\frac{(Q_2 - 2\Delta)}{\left(1+\frac{Q_2^2}{4}\right)} \pm \frac{1}{6}\frac{\sqrt{4-3Q_2^2}}{\left(1+\frac{Q_2^2}{4}\right)}\sqrt{\Theta} \quad (23)$$

$$Y_\pm = \frac{1}{3}\left(2\Delta \pm \sqrt{\Delta^2 - 3}\right) \quad (24)$$

which reduces to the well-known equation (24) when $Q_2=0$. In Fig. 2 (b), $X_\pm$ and $X_{2PA\pm}$ represent the value of normalized input pump power in absence and in presence of 2PA, respectively at which the bistability starts to occur (saddle nodes). Thus, it is to be noted from Fig. 2 (b), if there is no 2PA, both the curves (red curve) for $X_+$ ($X_-$, blue curve) and the cyan curve for $X_{2PA+}$ ($X_{2PA-}$, magenta) initiate from the same point. For $Q_2=0.2$, when the detuning, $\Delta$ is 2.216, both the values $X_+$ and $X_-$ merge into 2.7762 and indicates that if the $Q_2$ is large, bistability does not occur. As an example that the 2PA increases the bistability threshold, we present Fig. 2 (c)–2 (f).

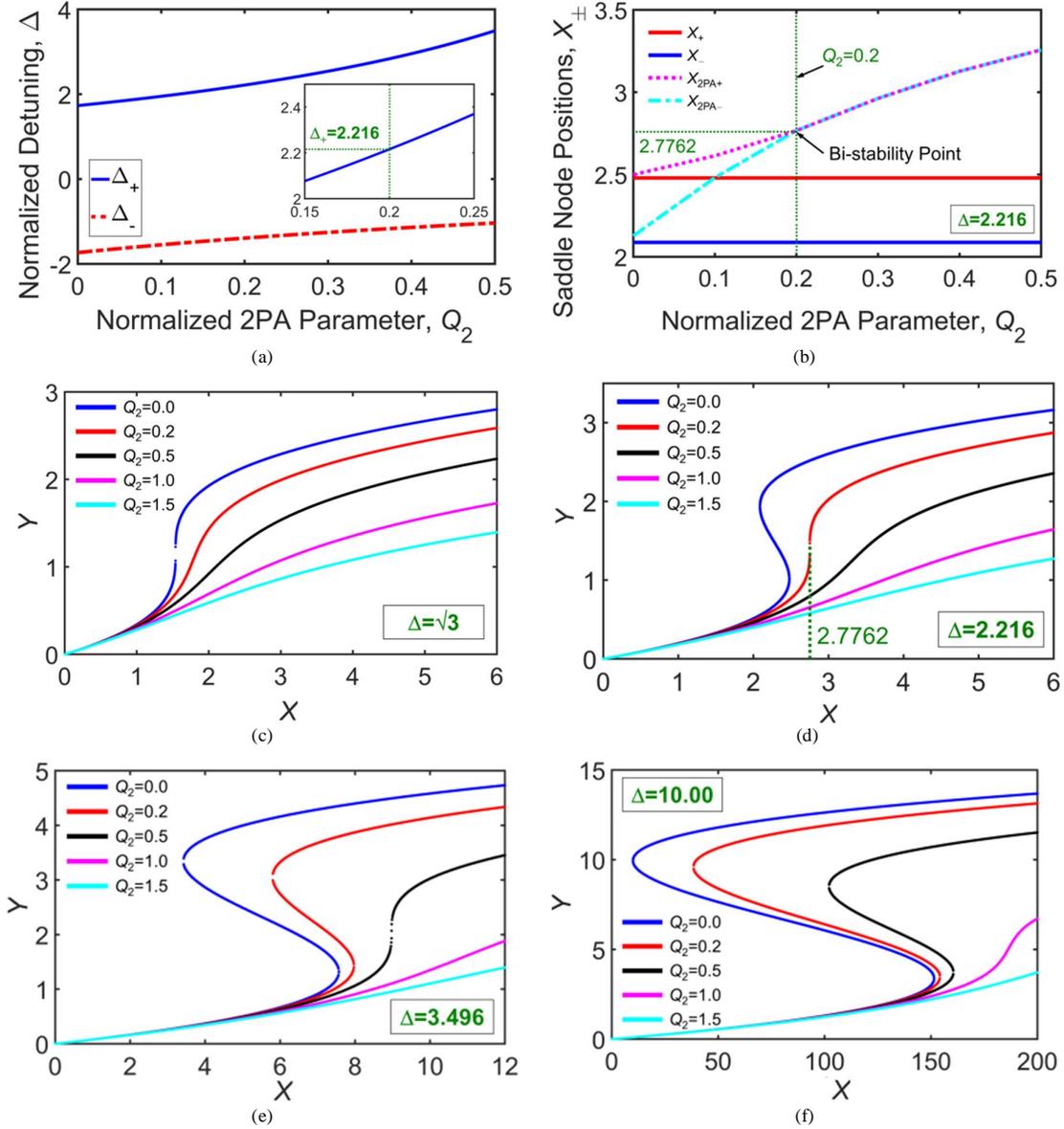

Fig. 2. (a) Normalized detuning $\Delta$ vs. normalized 2PA coefficient, $Q_2$, (b) Saddle node positions, $X_\pm$ and $X_{2PA\pm}$ with $Q_2$ when $\Delta$ is fixed at 2.216. Intra-cavity power $Y$ with the change in input pump power, $X$ (i.e. the bistability curve) for (c) $\Delta=\sqrt{3}$, (d) $\Delta=2.216$, (e) $\Delta=3.496$, and (f) $\Delta=10$, with a set of five different values of $Q_2$, in each case ($Q_2=0$, 0.2, 0.5, 1.0 and 1.5). From Fig. 2 (c), (d), (e) and (f) it is evident that bi-stable behavior initiates at $\Delta=\sqrt{3}$, 2.216, and 3.496 when the $Q_2$ is 0, 0.2 and 0.5, respectively, whereas, for $Q_2>2/\sqrt{3}$, bistability does not occur even if $\Delta$ is as high as 10.0 ( (a) and (f)) with positive and realistic values of $X$ and $Y$.

From Fig. 2 (c), 2 (d) and 2 (e) it can be seen that if there is no 2PA, the bistability starts to occur at $\Delta=\sqrt{3}$. However, if the 2PA coefficient $Q_2$ is 0.2 and 0.5, the minimum pump detuning that initiates the bistability are 2.216 and 3.496, respectively. Also, it can easily be found that, for any of the two roots ($\Delta_+$ or $\Delta_-$) to be positive, the condition given by equation (25) has to be satisfied:

$$Q_2 \leq \frac{2}{\sqrt{3}} \text{ or } Q_2 \geq 2\sqrt{3} \tag{25}$$

Earlier, it has been predicted numerically that after a certain maximum value of 2PA coefficient bistability may cease to exist [62]. In contrast, our theoretical model predicts that it is possible to observe bistability at $Q_2 = 2\sqrt{3}$ with a very small detuning, $\Delta$ (even less than $\sqrt{3}$ ) (as shown in Fig. 3 (a)–3 (d)) however to realize it, input pump power $X$ and intra-cavity power $Y$ both have to be negative, which cannot be possible. Therefore, we restrict our analysis for 2PA coefficient and discard other possibilities. It should also be noted that although optical bistability is a necessary condition for the formation of the optical cavity soliton but it is not a sufficient condition [62] to generate the cavity soliton.

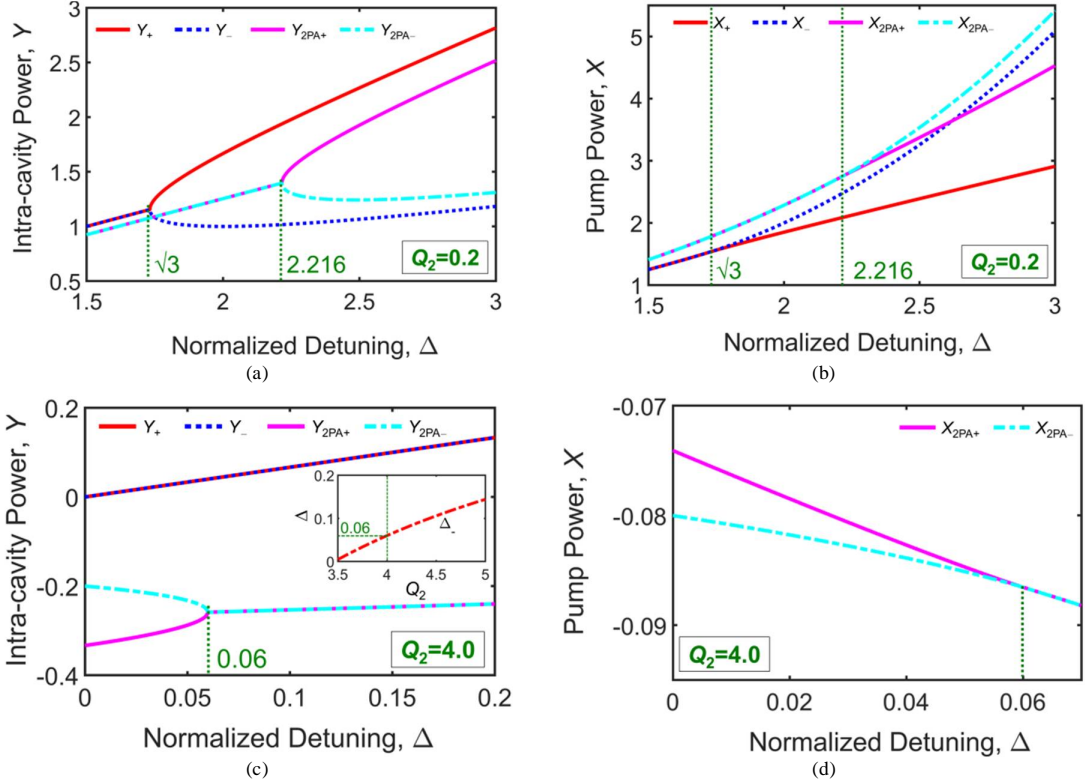

Fig. 3. In this figure the possibility of bistability for 2PA coefficient at different values (<2/√3) and (>2√3) are shown. (a) Intra-cavity power, $Y$ with respect to normalized detuning, $\Delta$ when the $Q_2$ is 0.2, (b) Input pump power $X$ vs. $\Delta$, at $Q_2$ is 0.2, (c) Intra-cavity power, $Y$ with respect to normalized detuning, $\Delta$ at $Q_2$=4.0, (d) Input pump power $X$ vs. $\Delta$, at $Q_2$=4.0. It is seen that bistability occurs at a positive value of $Y$ and $X$ if $Q_2$=0.2 (<2/√3), whereas, bistability can be obtained for a very small value of $\Delta$, with negative values of $X$ and $Y$ when $\Delta$=4 (>2√3).

### B. Kerr-tilt with Nonlinear Losses

Due to $\chi^{(3)}$ nonlinearity the maximum resonance peak occurs at normalized pump detuning $\Delta=Y=X$ which can be shown through the Kerr-tilt of the resonance peaks [61]; However, the $\Delta$ ($=Y$) for which the maximum intra-cavity power can be obtained in presence of one of the nonlinear losses (either 2PA, 3PA or 4PA) neglecting the FCA and FCD are the real root of equation (26).

$$\frac{Q_n^2}{n^2}Y^{2n-1} + \frac{2}{n}Q_n Y^n + Y - X = 0 \tag{26}$$

Thus the real roots of equation (26) yield the maximum value of $Y$ i.e. $Y_{max}$ for a particular normalized pump power, $X$ in presence of 2PA ($n$=2). The ratio of normalized intra-cavity power with respect to the normalized pump power (i.e. $Y_{max}/X$) and the corresponding normalized detuning, $\Delta$ at which the maxima occur are plotted in Fig. 4 (a) and 4 (b), respectively with varying $Q_2$ (from 0 to 0.5). Figs. 4 (c) and 4 (d) show the Kerr-tilt for three distinct values of $Q_2$ (0, 0.2, and 0.5) for two different input pump powers ($X$=1 and 1.5). It is apparent that higher the input pump power $X$, greater is the slope of the Kerr-tilt which means one requires large external pump detuning to obtain maximum intra-cavity power. In addition, if the 2PA is present, with the increase in input pump power, intra-cavity power decreases as the nonlinear absorption is more prominent at comparatively high input power.

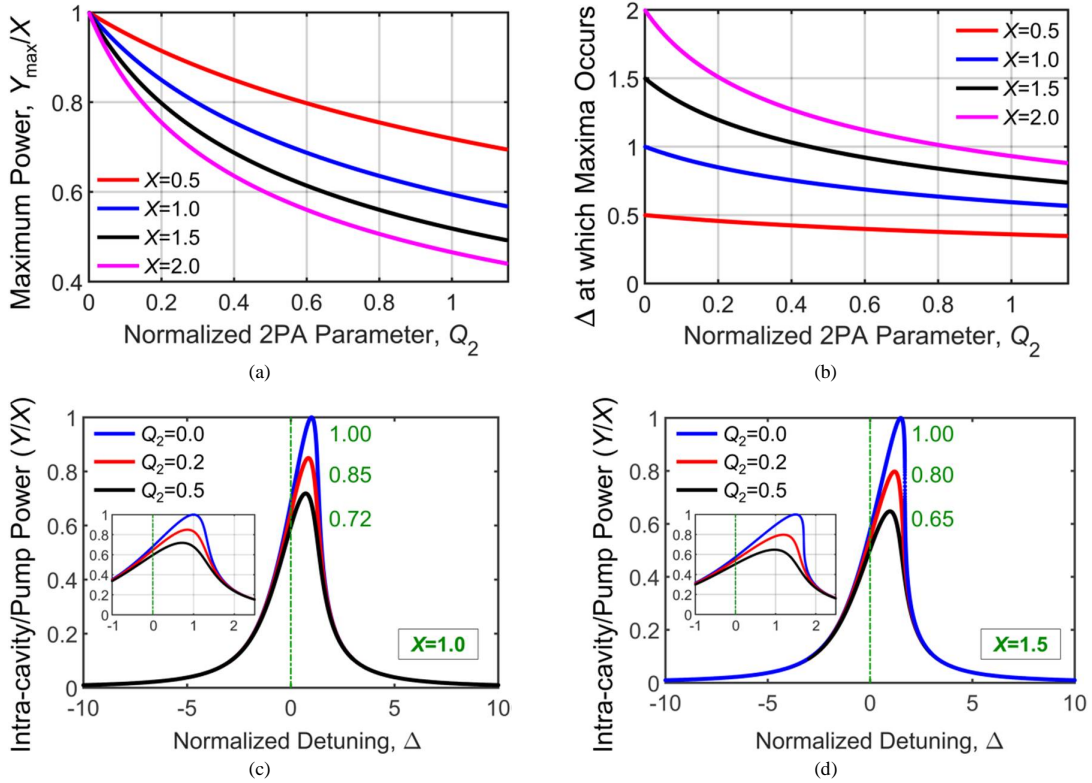

Fig. 4 (a) Ratio of maximum normalized intra-cavity power, $Y_{max}$ and the normalized pump power $X$ with various 2PA coefficients, $Q_2$, (b) Normalized detuning, $\Delta$ with 2PA coefficient, $Q_2$ where the maxima occur. Kerr tilt with different 2PA coefficients, when normalized input power is: (c) $X=1$, (d) $X=1.5$.

## C. Bistability & Kerr-tilt in Presence of FCA-FCD: Observation of Reverse (or Negative Kerr-tilt)

The previous section has provided a detailed quantitative analysis of the shift in Kerr-tilt as well as the reduction in the intra-cavity output power, $Y$ in presence of 2PA. In this section, we discuss the optical bistability ($Y$ vs. $X$) and Kerr-tilt ($Y/X$ vs. $\Delta$) in presence of FCA-FCD. At first, we start with arbitrary values of FCA and FCD-coefficients. From Figs. 5 (a) and 5 (b), it is evident that FCD introduces cavity detuning which manifests itself through the reverse Kerr-tilt. In practice, reverse Kerr-tilt shown in Fig. 5 (b) indicates that with suitable FCD-induced cavity detuning one can obtain maximum intra-cavity power even without any pump detuning (pink curve). For a fixed value of FCA-FCD coefficient the reverse tilt increases with the increase in the input pump-power, $X$ as depicted for two different values of FCA-FCD coefficients ($C_2=K_2=2$ and $C_2=K_2=5$) in Fig. 5 (c) and (d), respectively. $Y$ becomes multivalued with large detuning and pump power. In the next section we show that FCD-induced detuning, $\Delta_{FCD}$ depends upon the intra-cavity power. We observe that the nonlinear cavity detuning introduced only by 3PA at a very low pump power does not play any vital role in comb dynamics. However, along with reverse Kerr-tilt, in subsequent sections, it is shown that nonlinear losses especially 2PA, FCA, and FCD play crucial roles in determining the parameter spaces for MI, chaos formation and stable frequency-comb generations.

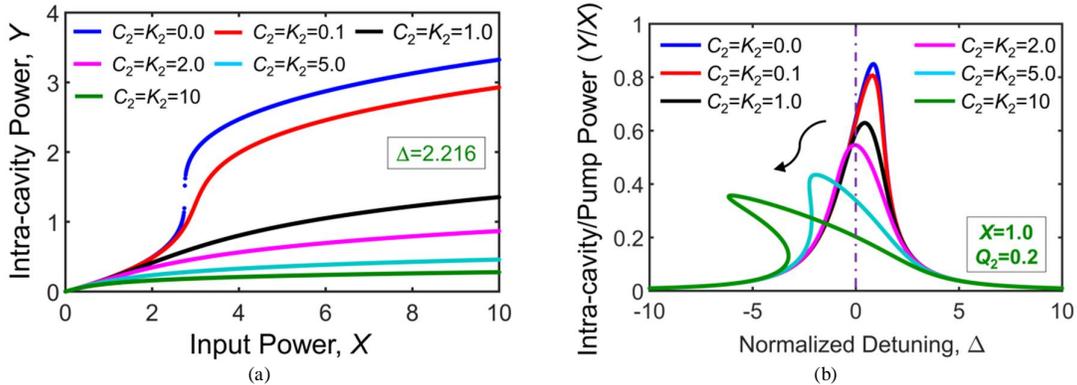

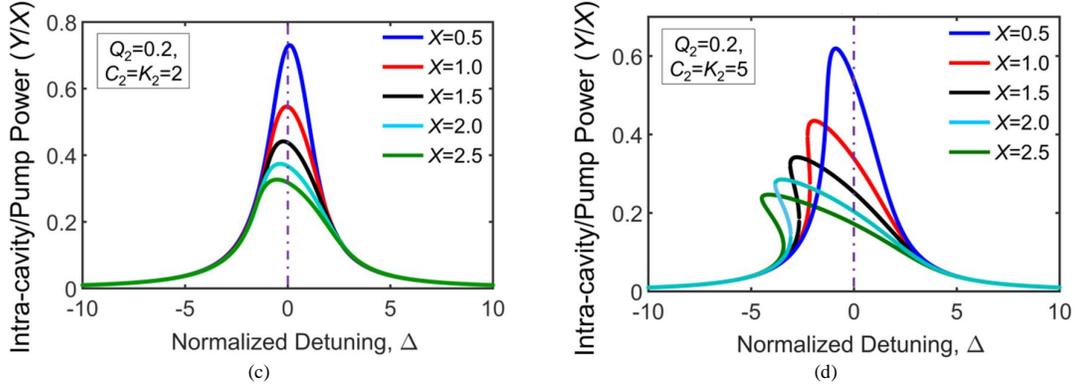

Fig. 5. (a) Bistability curve ($Y$ vs. $X$) with $\Delta$=2.216 and (b) Kerr-tilt ($Y/X$ vs. $\Delta$) at $X$=1, for a set of six different values of 2PA induced FCA-FCD ($C_2$ and $K_2$, respectively). Dependence of Kerr tilt ($Y/X$ vs. $\Delta$) on input pump power, $X$ when (c) $C_2$=$K_2$=2 and (d) $C_2$=$K_2$=5. For all the cases, (a)-(d) the 2PA coefficient, $Q_2$ is taken as 0.2.

### D. FCA-FCD induced cavity detuning

To explain the reverse Kerr-tilt, analytically, we further modify the normalized LLE (7) as:

$$\frac{\partial u}{\partial \xi} = -(1+j\Delta_{eff})u - \frac{1}{2}\phi_c u - j\frac{s}{2}\frac{\partial^2 u}{\partial \tau^2} + \sum_{k\geq 3}\frac{d_k}{k!}\frac{\partial^k u}{\partial \tau^k} + j|u|^2 u - \frac{Q_n}{n}|u|^{2(n-1)}u + S \quad (27)$$

$$\Delta_{eff} = \Delta + \Delta_{FCD} = \Delta + \frac{K\phi_c}{2} \quad (28)$$

where the effective cavity detuning, $\Delta_{eff}$ can be defined as the sum of the external pump detuning, $\Delta$ and the FCD-induced detuning, $\Delta_{FCD}$. It should be noted that this $\Delta_{eff}$ changes over round-trips and can have multiple values. Effective detuning, $\Delta_{eff}$ (solid red curve) and FCD-induced detuning, $\Delta_{FCD}$ (solid green-curve), with the variation in pump detuning when $C_2$=$K_2$=2, and $C_2$=$K_2$=5 at input pump power, $X$=1 are plotted in Figs. 6 (a) and (b), respectively. In both the cases, input pump detuning is also plotted (solid blue-curve) to indicate the reference level in absence of FCA-FCD. Generation of stable frequency comb even without external pump detuning with suitable initial pump power has been demonstrated in [51]. We also plot the ($Y/X$) with the change in $\Delta_{eff}$ instead of $\Delta$ in Figs. 6 (c) and (d) with two different values of $X$. As ($Y/X$) is now plotted with $\Delta_{eff}$, no reverse tilt can be observed. In steady state, the effective detuning, $\Delta_{eff}$ can be written as:

$$\Delta_{eff}^{(Steady\text{-}state)} = \Delta + \frac{KC_n|u_0|^{2n}}{2} = \Delta + \frac{KC_n Y_0^n}{2} \quad (29)$$

Where $u_0$ is the steady-state amplitude and the $Y_0$ is the steady-state intra-cavity power. The steady-state maximum FCD-induced cavity detuning for 2PA is plotted in Fig. 6 (e) whereas for 3PA and 4PA are plotted in Fig. 6 (f). It can be seen that the $\Delta_{FCD}$ is much larger in case of 2PA in comparison with 3PA and 4PA. To obtain steady-state maximum FCD induced cavity detuning for 2PA, 3PA, and 4PA (as shown in Figs. 6 (e) and 6 (f)), we have considered realistic values of FCA-FCD coefficients from the Table 2.

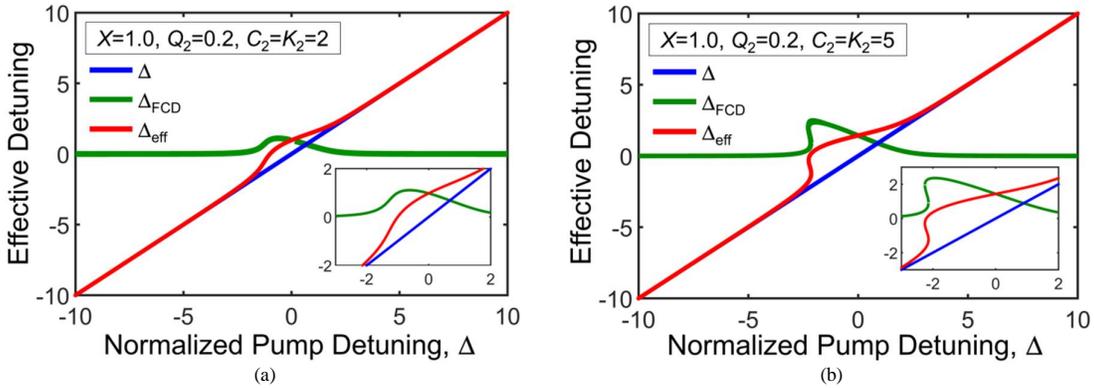

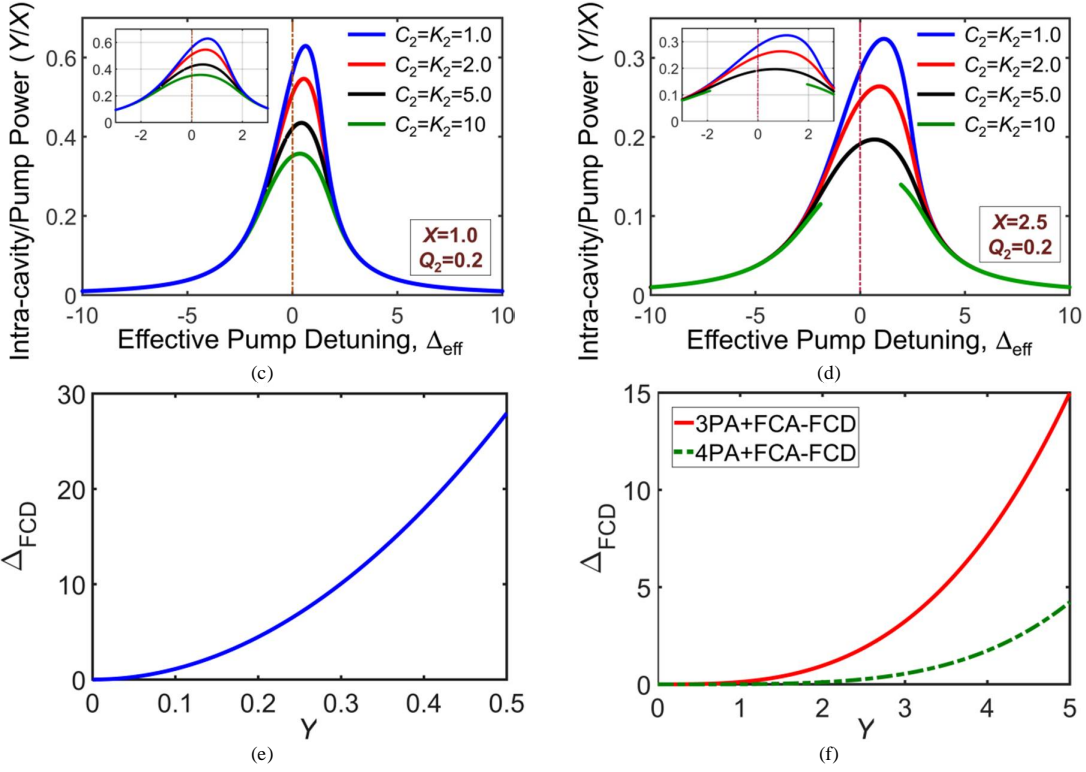

Fig. 6. Effective detuning, $\Delta_{eff}$ (red-curve) and FCD-induced detuning, $\Delta_{FCD}$ (green-curve), with the variation in pump detuning when (a) $C_2=K_2=2$, and (b) $C_2=K_2=5$ at input pump power, $X=1$ in both the cases. Input pump detuning is also plotted (blue-curve) both in (a) and (b) to indicate the reference level in absence of FCA-FCD. Same Kerr-tilt ($Y/X$) with the change in effective detuning, $\Delta_{eff}$, instead of pump detuning, $\Delta$ for two different pump powers: (c), $X=1$ and (d) $X=2.5$. FCD induced steady-state cavity detuning, $\Delta_{FCD}$ with the intra-cavity power, $Y$ when (e) the free-carriers are generated due to the 2PA (blue solid curve), and (f) in presence of 3PA (red solid curve) and 4PA (green dotted curve). $\Delta_{FCD}$ can possess multiple values if FCA-FCD coefficients are high. Fig. (c) and (d) show that if ($Y/X$) is plotted against $\Delta_{eff}$ instead of $\Delta$, no reverse Kerr-tilt can be observed. Finally, it can be seen from Fig. (e) and (f) that the $\Delta_{FCD}$ for the generated free-carriers due to 2PA is much larger compared to that of due to the 3PA and 4PA.

## V. Linear Stability Analysis

In this section, we perform the linear stability analysis of the stationary CW solutions ($u_0$ and $\phi_0$) of free-carrier driven LLE. The evolution of normalized carrier density and signal amplitude due to the spatiotemporal perturbations is given by equations (30) and (31), respectively while nonlinear losses (any of the 2PA/TPA, 3PA, or 4PA) along with corresponding FCA-FCD. In each case, we find the MI-gain, $\lambda$. We also show the dependence of MI gain and bandwidth on normalized detuning, $\Delta$, signal power ($Y$, which is dependent upon pump power, $X$) and higher-order dispersion. Finally, we provide a general expression of MI-gain for multi-photon absorptions. In our analytical expressions superscript ($n$) stands for $n$PA.

$$\phi_c^{(n)}(\xi,\tau) = \phi_0^{(n)}(\xi) + \phi_+^{(n)}(\xi)e^{j\Omega\tau} + \phi_-^{(n)}(\xi)e^{-j\Omega\tau} \tag{30}$$

$$u^{(n)}(\xi,\tau) = u_0^{(n)}(\xi) + u_+^{(n)}(\xi)e^{j\Omega\tau} + u_-^{(n)}(\xi)e^{-j\Omega\tau} \tag{31}$$

Thus, the general expression for homogeneous solutions for the normalized free-carrier and the intra-cavity field amplitude ($\phi_0^{(n)}$, $u_0^{(n)}$, respectively) can be given as:

$$\phi_0^{(n)} = \theta_{c_n}\tau_c \left|u_0^{(n)}\right|^{2n} = C_n \left|u_0^{(n)}\right|^{2n} \tag{32}$$

$$\frac{\partial u_0^{(n)}}{\partial \xi} = -(1+j\Delta)u_0^{(n)} - \frac{1}{2}(1+jK)\theta_{c_n}\tau_c \left|u_0^{(n)}\right|^{2n} u_0^{(n)} + j\left|u_0^{(n)}\right|^2 u_0^{(n)} - \frac{Q_n}{n}\left|u_0^{(n)}\right|^{2(n-1)} u_0^{(n)} + S \tag{33}$$

Whereas, the perturbation in free-carriers ($\phi_+^{(n)}$ and $\phi_-^{(n)}$) can be given by:

$$\phi_+^{(n)} = \frac{n\theta_{c_n}\tau_c \left|u_0\right|^{2(n-1)}\left[u_0^* u_+ + u_0 u_-^*\right]}{(1+j\Omega\tau_c)} = \phi_-^{(n)*} \tag{34}$$

We assume, $u_+=C.\exp(\lambda.\xi)$ where $C$ is a constant. Finally, the MI-gain can be found by solving the quadratic equation of $\lambda$ (35),

$$\left( \frac{\{R^{(n)}\}^2 + \{I^{(n)}\}^2}{\{\alpha^{(n)}\}^2 + \{\beta^{(n)}\}^2} - Y^2 \right) = 0 \tag{35}$$

Here, the LLE is truncated up to the 3rd order dispersion. $R^{(n)}$, $I^{(n)}$, $\alpha^{(n)}$, and $\beta^{(n)}$ are given as,

$$R^{(n)}(\lambda) = -\left[ (1+\lambda) + \frac{C_n}{2}\left\{\frac{(1+K\Omega\tau_c)}{(1+\Omega^2\tau_c^2)}\right\}nY^n + \frac{C_n}{2}Y^n + Q_nY^{(n-1)} \right] \tag{36}$$

$$I^{(n)} = -\left[ \Delta + \frac{C_n}{2}\left\{\frac{(K-\Omega\tau_c)}{(1+\Omega^2\tau_c^2)}\right\}nY^n + \frac{KC_n}{2}Y^n - \frac{s}{2}\Omega^2 + \frac{\Omega^3}{6}d_3 - 2Y \right] \tag{37}$$

$$\alpha^{(n)} = \left[ \frac{C_n}{2}\left\{\frac{(1+K\Omega\tau_c)}{(1+\Omega^2\tau_c^2)}\right\}nY^{(n-1)} + \frac{Q_n}{n}(n-1)Y^{(n-2)} \right] \tag{38}$$

$$\beta^{(n)} = \left[ \frac{C_n}{2}\left\{\frac{(K-\Omega\tau_c)}{(1+\Omega^2\tau_c^2)}\right\}nY^{(n-1)} - 1 \right] \tag{39}$$

Equation (35) can also be simplified in a more convenient form given by equation (40),

$$\left[ \lambda^2 + 2\varepsilon^{(n)}\lambda + \left(\varepsilon^{(n)}\right)^2 + \left(I^{(n)}\right)^2 - \left\{\left(\alpha^{(n)}\right)^2 + \left(\beta^{(n)}\right)^2\right\}Y^2 \right] = 0 \tag{40}$$

Where $\varepsilon^{(n)}$ is expressed as,

$$\varepsilon^{(n)} = \left[ 1 + \frac{C_n}{2}\left\{\frac{(1+K\Omega\tau_c)}{(1+\Omega^2\tau_c^2)}\right\}nY^n + \frac{C_n}{2}Y^n + Q_nY^{(n-1)} \right] \tag{41}$$

Note that, in absence of all nonlinear losses (multi-photon absorption, FCA, FCD) and the higher order dispersion terms, $\varepsilon^{(n)}$, $I^{(n)}$, $\alpha^{(n)}$ and $\beta^{(n)}$ becomes 1, $-(\Delta - s\Omega^2/2 - 2Y)$, 0 and 1, respectively such that $\lambda$ reduces to the well-known expression (42) for MI-gain [57-58]:

$$\lambda_\pm(\Omega) = -1 \pm \sqrt{4Y\left(\Delta - \frac{s\Omega^2}{2}\right) - \left(\Delta - \frac{s\Omega^2}{2}\right)^2 - 3Y^2} \tag{42}$$

In the subsequent sections, we plot the MI-gain with respect to normalized frequency $\Omega$, normalized detuning $\Delta$ as well as normalized intra-cavity power $Y$ both for anomalous ($s=-1$) and normal ($s=1$) dispersions in absence and presence of all nonlinear losses and compare our results with results published in Ref. [57].

### A. MI in Normal Dispersion Region (s=1)

It is known that unlike straight waveguide or optical fiber, MI can occur in synchronously driven optical cavities or in ring lasers even if the system is pumped in normal dispersion regime [57]. As a result, a stable stationary train of pulses can be generated in the cavity irrespective of the sign of the dispersion.

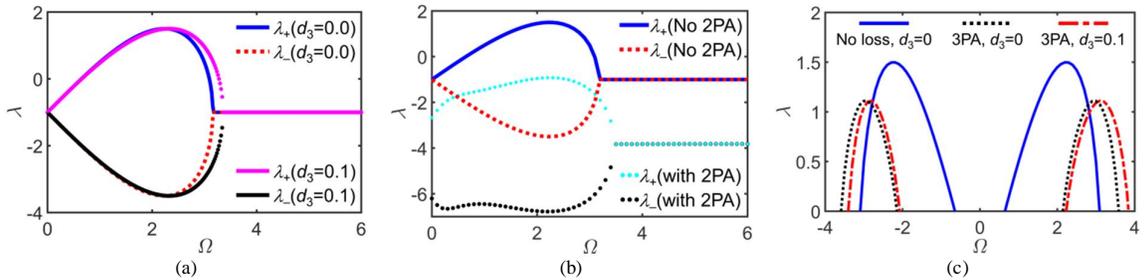

Fig. 7. (a) Positive (when $\Omega>0$) MI-gain lobes $\lambda_\pm$ in absence and presence of third order dispersion ($d_3=0.1$), (b) $\lambda_\pm$ in absence and presence of 2PA, (c) $\lambda_\pm$ in presence and absence of 3PA. 3rd order dispersion induces asymmetry in the gain lobe with respect to $\Omega=0$. Simulation parameters are: $s=1$, $Y=2.5$, $\Delta=7.5$, $Q_2=1$, $K_2=0.05$, $C_2=0.1$, $\theta_{c2}=0.05$, $\theta_{c3}=6.31\times10^{-7}$, $Q_3=8.73\times10^{-4}$, $C_3=0.049$, $K_3=4.9$.

In this section, we discuss the dependency of MI-gain $\lambda$ on $\Omega$ and $\Delta$ (at $Y=2.5$) as well as on $\Omega$ and $Y$ (at $\Delta=7.5$), respectively in normal dispersion region ($s=1$). At first, we plot MI-gain, $\lambda$ with respect to normalized frequency $\Omega$ in Fig. 7 for normal ($s=1$) dispersion regime. Figs. 7 (a) and 7 (b) show that a non-zero third-order dispersion coefficient, $d_3$ (=0.1) can enhance the MI-bandwidth whereas, the MI-gain in presence of 2PA with FCA and FCD is insufficient to create any sidebands. In addition, $d_3$ induces the asymmetry in MI-gain-lobe with respect to $\Omega=0$ (Fig. 7 (c)). The values of $n$PA, FCA-FCD coefficients are taken from Table 2.

Figs. 8 (a) and 8 (b) depict the dependency of MI-gain $\lambda$ on $\Omega$ and $\Delta$ (at $Y=2.5$) as well as on $\Omega$ and $Y$ (at $\Delta=7.5$), respectively in normal dispersion region while all the nonlinear losses are ignored.

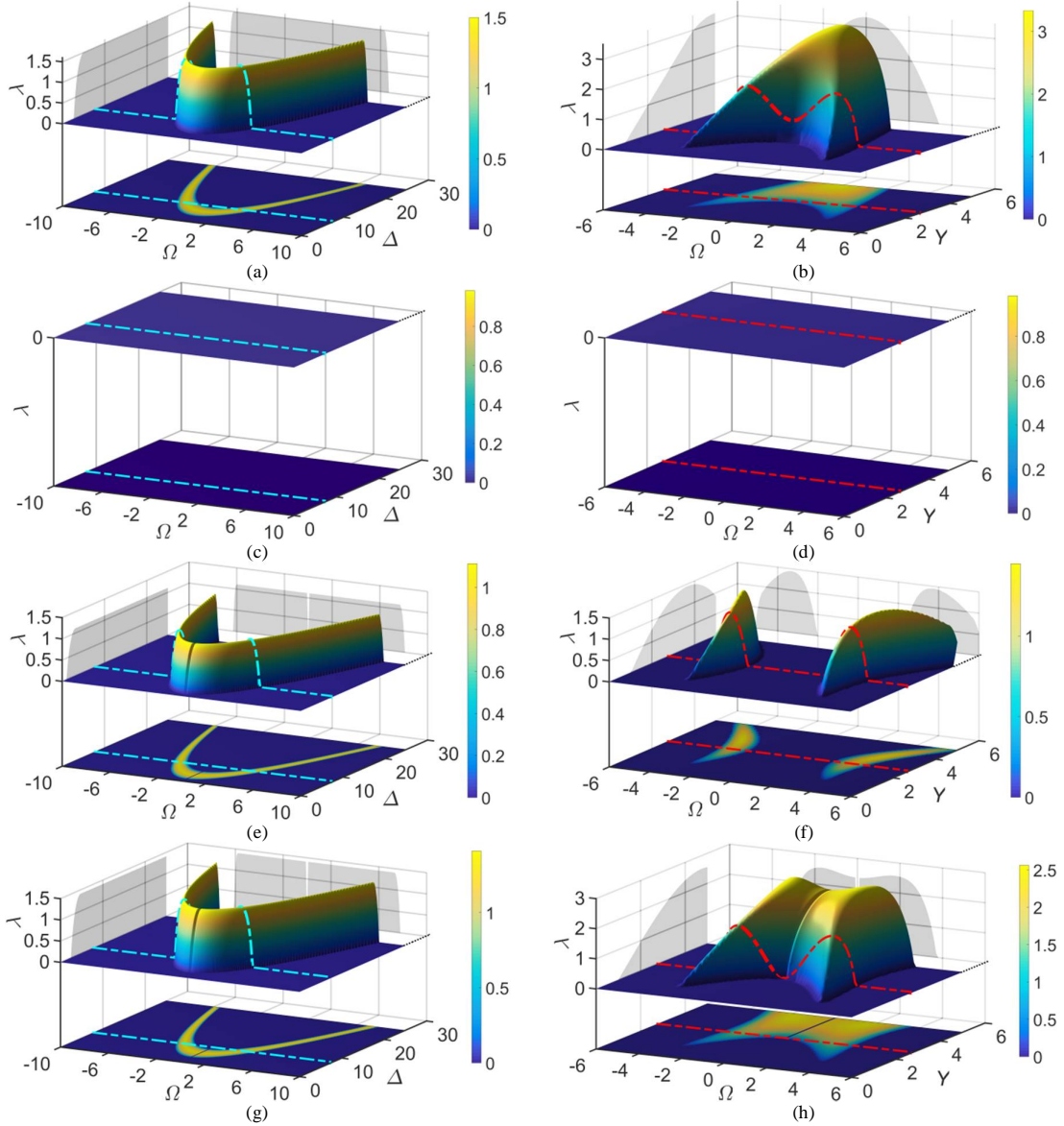

Fig. 8. Real part of the MI-gain $\lambda$ with (a) $\Omega$ and $\Delta$ (at $Y=2.5$), with (b) $\Omega$ and $Y$ (at $\Delta=7.5$) in absence of all nonlinear losses. Variation of $\lambda$ with (c) $\Omega$ and $\Delta$ (at $Y=2.5$), (d) $\Omega$ and $Y$ (at $\Delta=7.5$) while 2PA, FCA-FCD are present. Change in $\lambda$ with the change in (e) $\Omega$ and $\Delta$ (at $Y=2.5$), and (f) $\Omega$ and $Y$ (at $\Delta=7.5$) when 3PA, FCA-FCD are present. $\lambda$ with (g) $\Omega$ and $\Delta$ (at $Y=2.5$), and (h) $\Omega$ and $Y$ (at $\Delta=7.5$) in presence of 4PA, FCA-FCD. For all the cases, $3^{rd}$ order dispersion ($d_3$=0.1) is taken into consideration. Other simulation parameters are: $s=+1$ (normal dispersion), $\theta_{c2} = 0.0005$, $Q_2 = 0.93$, $C_2 = 29.81$, $K_2 = 7.5$ (at $\lambda_0 \sim 1.56\mu m$); $\theta_{c3} = 6.31 \times 10^{-7}$, $Q_3 = 8.73 \times 10^{-4}$, $C_3 = 0.049$, $K_3 = 4.9$ (at $\lambda_0 \sim 2.4\mu m$); $\theta_{c4} = 6.018 \times 10^{-8}$, $Q_4 = 6.16 \times 10^{-6}$, $C_4 = 4.66 \times 10^{-3}$, $K_4 = 2.9$ (at $\lambda_0 \sim 4.0\mu m$) (as given in table-II). The cyan and magenta lines overlaid on the 2-D plots are indicating the corresponding values of $\lambda$ for $\Delta = 7.5$ and $Y = 2.5$ respectively.

To produce rest of curves (Fig 8 (c)-8 (h)) we use the realistic values of $n$PA, FCA-FCD coefficients for silicon waveguides from Table 2. For silicon, in presence of 2PA no parametric gain lobe has been observed (Fig. 8 (c) and 8 (d)). We plot the $\lambda$ in presence of 3PA and 4PA with the corresponding FCA-FCD in Fig. 8 (e)-8 (h). The cyan and magenta lines are drawn on the 2-D plots to indicate the corresponding values of $\lambda$ for $\Delta = 7.5$ and $Y = 2.5$, respectively [57]. Projection of the 2-D plot on each axis through the shadow-plot helps the readers to anticipate the approximate values of the MI-gain with respect to different parameters.

### B. *MI in Anomalous Dispersion Region (s=−1)*

Figs. 9 (a) and 9 (b) show the dependency of MI-gain $\lambda$ on $\Omega$ and $\Delta$ (at $Y=2.5$) as well as on $\Omega$ and $Y$ (at $\Delta=7.5$), respectively in anomalous dispersion region ($s=-1$) of operation while all the nonlinear losses are negligible. To produce rest of curves (Fig 9 (c)-9 (h)) we use the realistic values of $n$PA, FCA-FCD coefficients for silicon waveguides from Table 2. There is no parametric gain in presence of 2PA which is shown in Figs. 9 (c) and 9 (d). Similar to the Fig. 8, we plot the $\lambda$ in presence of 3PA and 4PA with the corresponding FCA-FCD in Fig. 9 (e)-9 (h).

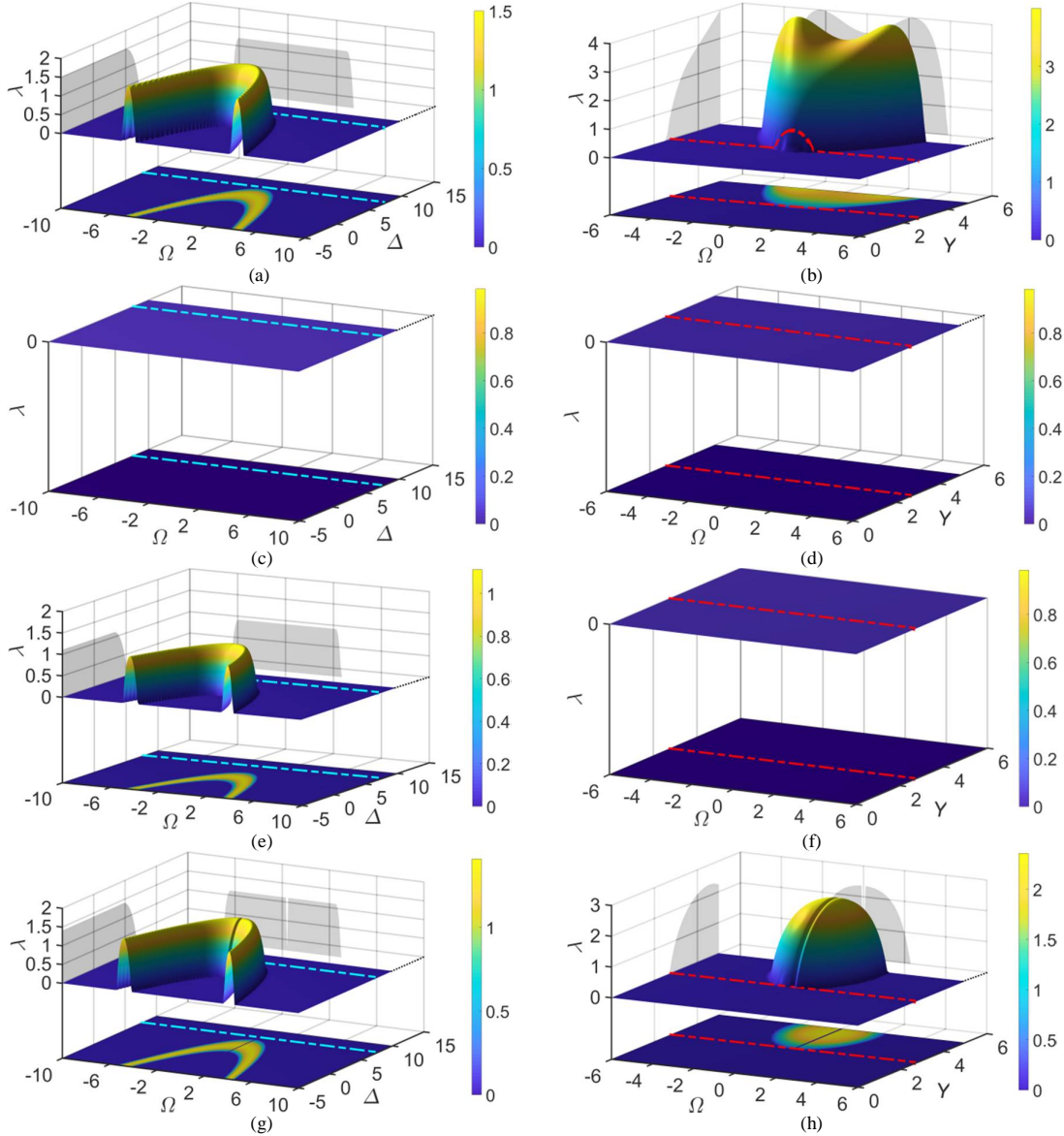

Fig. 9. Real part of the MI-gain $\lambda$ with (a) $\Omega$ and $\Delta$ (at $Y=2.5$), with (b) $\Omega$ and $Y$ (at $\Delta=7.5$) in absence of all nonlinear losses. Variation of $\lambda$ with (c) $\Omega$ and $\Delta$ (at $Y=2.5$), (d) $\Omega$ and $Y$ (at $\Delta=7.5$) while 2PA, FCA-FCD are present. Change in $\lambda$ with the change in (e) $\Omega$ and $\Delta$ (at $Y=2.5$), and (f) $\Omega$ and $Y$ (at $\Delta=7.5$) when 3PA, FCA-FCD are present. $\lambda$ with (g) $\Omega$ and $\Delta$ (at $Y=2.5$), and (h) $\Omega$ and $Y$ (at $\Delta=7.5$) in presence of 4PA, FCA-FCD. For all the cases, $3^{rd}$ order dispersion ($d_3=0.1$) is taken into consideration. Other simulation parameters are: $s=-1$ (anomalous dispersion), $\theta_{c2} = 0.0005$, $Q_2 = 0.93$, $C_2 = 29.81$, $K_2 = 7.5$ (at $\lambda_0 \sim 1.56 \mu m$); $\theta_{c3} = 6.31 \times 10^{-7}$, $Q_3 = 8.73 \times 10^{-4}$, $C_3 = 0.049$, $K_3 = 4.9$ (at $\lambda_0 \sim 2.4 \mu m$); $\theta_{c4} = 6.018 \times 10^{-8}$, $Q_4 = 6.16 \times 10^{-6}$, $C_4 = 4.66 \times 10^{-3}$, $K_4 = 2.9$ (at $\lambda_0 \sim 4.0 \mu m$) (as given in table-II). The cyan and magenta lines overlaid on the 2-D plots are indicating the corresponding values of $\lambda$ for $\Delta = 7.5$ and $Y = 2.5$ respectively.

## C. Effect of FCA-FCD on MI-growth

The impact of FCA-FCD on MI-gain is demonstrated in Fig. 10. In case of 3PA, the principle mechanisms to inhibit the comb formation are 3PA-induced FCA and FCD [45]. To obtain parametric oscillation in presence of 3PA, either the generated free-carriers have to be swept away by suitable external-bias or the input pump power $X$ (consequently, $Y$) has to be sufficiently low. We plot the $\lambda$ with $\Omega$ and $\Delta$ in Fig. 10 (a) in absence of any nonlinear losses, whereas Fig. 10 (b) dictates the MI-gain in absence of FCA-FCD while 3PA is present. Fig. 10 (c) shows the MI-gain in presence of 3PA, FCA-FCD at $Y=2.5$. Similarly, when the intra-cavity power is almost quadrupled ($Y=10.5$) the corresponding MI-gain has been plotted in Fig. 10 (d)-10 (f) for three different cases as described earlier. It is conspicuous from Fig. 10 (c) and (f) that there is no parametric oscillation if FCA-FCD is present while the intra-cavity power is high ($Y=10.5$) however, MI-gain lobes exist for relatively low intra-cavity power ($Y=2.5$).

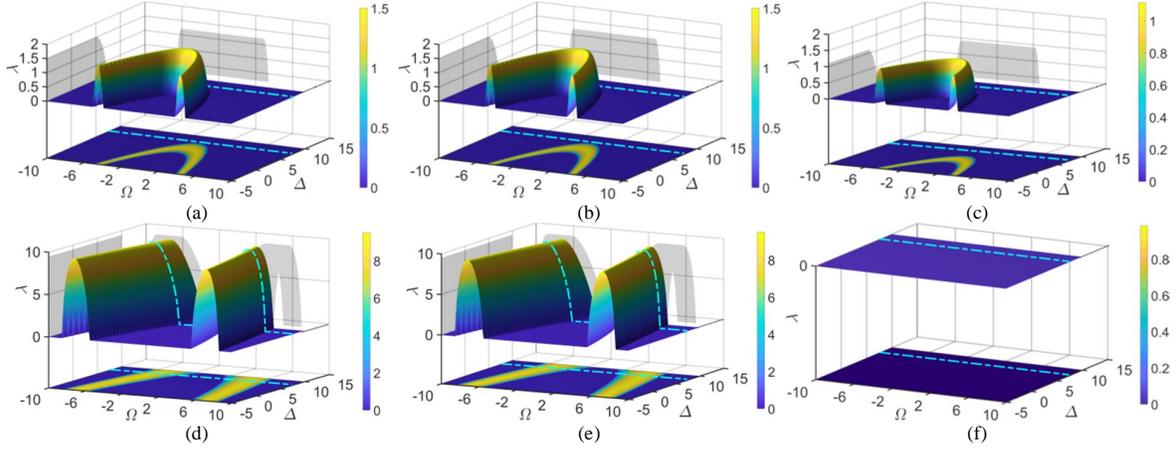

Fig. 10. MI-gain $\lambda$ with respect to $\Omega$ and $\Delta$ when $Y$ is fixed at 2.5 in (a) absence of all nonlinear losses, (b) presence of only 3PA and no FCA-FCD, (c) presence of 3PA, FCA-FCD. MI-gain $\lambda$ with respect to $\Omega$ and $\Delta$ when at $Y=10.5$ in (d) absence of all nonlinear losses, (e) presence of only 3PA and no FCA-FCD, (f) presence of 3PA, FCA-FCD. It is observed that if input power $X$ is high (high intra-cavity power, $Y$) parametric oscillation ceases to occur in presence of FCA-FCD along with 3PA.

To understand the significance of the FCA-FCD, we plot the MI-gain in Fig. 11 with respect to $\Omega$ and $Y$ at $\Delta=7.5$ when the $C_3$ and $K_3$, both the coefficients are (a) 2 times, (b) 5 times and (c) 10 times less than the original values listed in table-2.

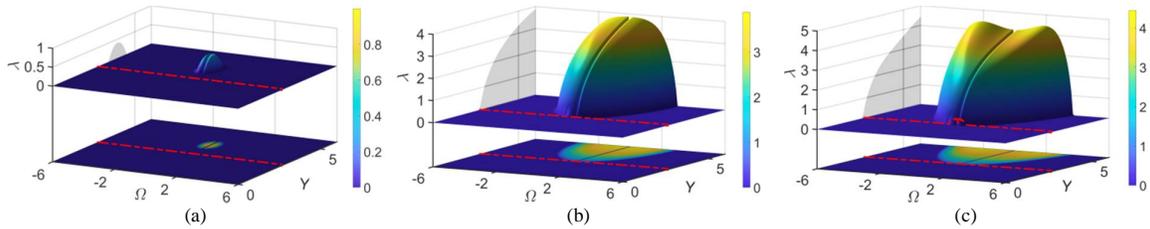

Fig.11 MI-gain $\lambda$ with respect to $\Omega$ and $Y$ when $\Delta$ is fixed at 7.5 when the $C_3$ and $K_3$ both the coefficients are (a) 2 times, (b) 5 times and (c) 10 times less than the original values listed in table-2.

## VI. Condition to obtain maximum MI-gain and Threshold Intensity

Starting from the well-known equation (42), the maximum gain is achieved when $\Delta k=0$, where $\Delta k$ is given by,

$$\Delta k = \left(\Delta - \frac{s}{2}\Omega^2 - 2Y\right) \tag{43}$$

where,

$$\lambda_{\pm}(\Omega) = -1 \pm \sqrt{Y^2 - (\Delta k)^2} \tag{44}$$

Thus, the maximum value of $\lambda$, $\lambda_{max}$ in absence of any nonlinear losses is ($Y-1$) and the steady state solution will always be stable if $Y<1$. However, in presence of nonlinear losses, the solution of $\lambda$ can be written as (45) instead of equation (44).

$$\lambda^{(n)} = -\varepsilon^{(n)} \pm \sqrt{\left\{\left(\alpha^{(n)}\right)^2 + \left(\beta^{(n)}\right)^2\right\} Y^2 - \left(I^{(n)}\right)^2} \quad (45)$$

Therefore, the maximum gain can be obtained if $I^{(n)}=0$. One can write the general expression of $\lambda_{max}$ in terms of $Y$, $\alpha^{(n)}$, $\beta^{(n)}$ and $\varepsilon^{(n)}$ as follows:

$$\lambda_{max}^{(n)} = Y\sqrt{\left\{\left(\alpha^{(n)}\right)^2 + \left(\beta^{(n)}\right)^2\right\}} - \varepsilon^{(n)} \quad (46)$$

If the FCA and FCD are neglected, the expression of maximum gain can be written as:

$$\lambda_{max}^{(n)} = Y\sqrt{\left(\frac{n-1}{n}\right)^2 Q_n^2 Y^{2(n-2)} + 1} - \left(1 + Q_n Y^{n-1}\right) \quad (47)$$

In this section, we also deduce the expression for the minimum intensity to initiate the MI. It can be shown easily that the steady-state solution will always be stable for,

$$Y \leq \frac{\varepsilon^{(n)}}{\sqrt{\left\{\left(\alpha^{(n)}\right)^2 + \left(\beta^{(n)}\right)^2\right\}}} \quad (48)$$

As an example, in presence of only 2PA, the value of $Y$ for which the steady-state solution of the LLE will always be stable can be given by,

$$Y \leq \frac{1}{\left(\sqrt{\frac{Q_2^2}{4}+1}\right) - Q_2} \quad (49)$$

Maximum MI-gain $\lambda_{max}$ is plotted in Fig. 12 (a) for four different values (0, 0.2, 0.5, and 0.9) of 2PA coefficients $Q_2$ in absence of FCA-FCD. It should be noted that 2PA coefficient for silicon at telecom wavelength, $Q_2$ is~0.9.

**VII. Range of Normalized Detuning to Obtain MI**

In this section we find the range of possible normalized detuning to initiate the MI and compare the results for different cases such as, while none of the nonlinear losses is present, only $n$PA is present, and $n$PA, FCA are present and all the nonlinear losses are present. The condition for threshold can be found equating $\lambda$ to 0. It is known that in absence of nonlinear losses,

$$\frac{s}{2}\Omega^2 = \left(\Delta - g_\pm\right) \quad (50)$$

When $g_\pm$ can be expressed in terms of $Y$,

$$g_\pm = 2Y \pm \sqrt{Y^2 - 1} \quad (51)$$

It is to be noted, in anomalous dispersion region, $s=-1$, for $\Omega$ to possess real solution, $\Delta$ must be $<g_+$ whereas for normal dispersion regime, the required detuning $\Delta$ should be$>g_-$ for MI to occur. When, the pump detuning lies in between $\in (g_-, g_+)$, i.e. $g_-<\Delta<g_+$, MI can be initiated for both the anomalous and normal dispersion regimes. Similarly, in presence of all the nonlinear losses the general expression to obtain the threshold condition can be found from equation (52),

$$\left\{I^{(n)}\right\}^2 = \left\{\left(\alpha^{(n)}\right)^2 + \left(\beta^{(n)}\right)^2\right\} Y^2 - \left\{\varepsilon^{(n)}\right\}^2 \quad (52)$$

However, the relation of MI Gain, $\Omega$ with $\Delta$ and output power $Y$ obtained from the equation (53)-(55) can be given by a 5$^{th}$ order polynomial of $\Omega$ which does not have any analytic solution. The following relation can be given as,

$$f(\Omega) = -\frac{d_3}{6}\tau_c^2\Omega^5 + \frac{s}{2}\tau_c^2\Omega^4 - \frac{d_3}{6}\Omega^3 + \left\{\frac{s}{2} - \left(\Delta - g_{1_\pm}\right)\tau_c^2\right\}\Omega^2 + \frac{\theta_{c_n}\tau_c^2}{2}nY^n\Omega = \left(\Delta - g'_\pm\right) \quad (53)$$

Where,

$$g_{1_\pm} = 2Y - \frac{\theta_{c_n}\tau_c}{2}KY^n \pm \sqrt{\left\{\left(\alpha^{(n)}\right)^2 + \left(\beta^{(n)}\right)^2\right\}Y^2 - \left(\varepsilon^{(n)}\right)^2} \quad (54)$$

$$g'_\pm = -(n+1)\frac{\theta_{c_n}\tau_c}{2}KY^n + 2Y \pm \sqrt{\left\{\left(\alpha^{(n)}\right)^2 + \left(\beta^{(n)}\right)^2\right\}Y^2 - \left(\varepsilon^{(n)}\right)^2} \quad (55)$$

Equation (53)-(55) can only be solved numerically. If the carrier-lifetime can be reduced by sweeping the carriers applied the external bias voltage across the device cross-section, which often can be done by forming a p-i-n junction across the waveguide cross-section then the square of $\tau_c$ term can be negligible. The situation is equivalent to the case when $n$PA is present while FCA and FCD are absent. In this case, neglecting the higher order dispersion terms ($d_k(k\geq3)=0$) the equations reduce to,

$$\frac{s}{2}\Omega^2 = \left(\Delta - g'_\pm\right) \tag{56}$$

Also in absence of FCA-FCD we can find the range of $\Delta$ analytically which is given by equations (57)-(58),

$$\frac{s}{2}\Omega^2 = \left(\Delta - g^{(n)}_\pm\right) \tag{57}$$

$$g^{(n)}_\pm = 2Y \pm \sqrt{\left(\left(\frac{n-1}{n}\right)^2 Q_n^2 Y^{2(n-2)} + 1\right)Y^2 - \left(1 + Q_n Y^{n-1}\right)^2} \tag{58}$$

In Fig. 12 (b) we plot the range of $\Delta$ both in absence of all nonlinear losses and in presence of only 2PA, for which the steady-state solution can be unstable. It is discussed earlier that in absence of all nonlinear losses the steady-state solution will be unstable in the range: $g_-<\Delta<g_+$ where $g_\pm$ is given by equation (51). From Fig. 12 (b) it can be seen clearly that with the gradual increase in the 2PA coefficient, the range of $\Delta$ for which MI can initiate becomes narrower.

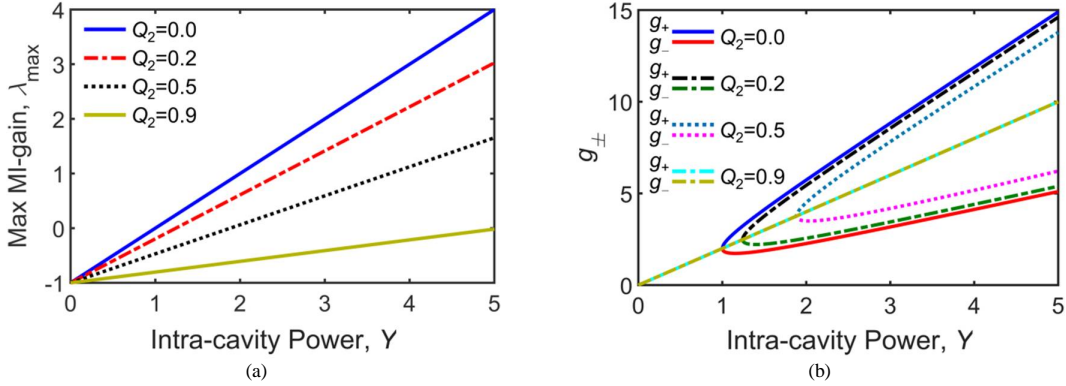

Fig. 12 (a) Maximum MI-gain with the change in intra-cavity power $Y$ with a set of four 2PA coefficients ($Q_2=0$, 0.2, 0.5, 0.9), (b) Range of $\Delta$ for which MI can initiate with different 2PA coefficients ($Q_2=0$, 0.2, 0.5, 0.9).

**VIII. Numerical Results**

Our theoretical predictions are confirmed numerically by solving the LLE (7) and the rate-equation (8) through split-step Fourier method. The normalized spectrum for three different scenarios is shown in Fig. 13, where solid blue curve explains the generation of FC in absence of any nonlinear losses, whereas the dashed red curve and dotted green curve correspond to the spectrum when 2PA and 3PA are present, respectively. It has been shown earlier that in presence of 3PA, MI can occur though the gain lobes are reduced significantly.

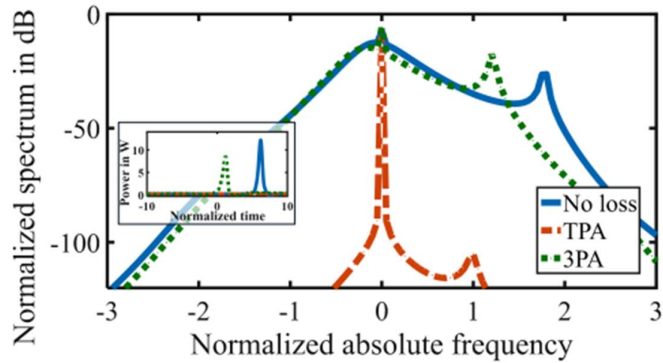

Fig. 13. Normalized spectrum of frequency-comb without any nonlinear losses (solid blue curve), with only 2PA (dashed red curve) and with 3PA (dotted green curve). The inset shows the final pulses in the time domain. Simulation parameters are taken from Table 2.

However, if corresponding FCA-FCD is present along with the 3PA, then the FCA-FCD can potentially inhibit the comb formation [45]. Effects of FCA-FCD can be reduced using a very low input pump as shown in Fig. 10. Similarly, in numerical simulations, it is observed that 2PA inhibits the comb formation, whereas if only 3PA is present the comb-bandwidth is reduced which is in congruence with the analytical formulations.

**IX. Conclusions**

To conclude our work, we derived analytical expressions of steady-state homogeneous solutions of free-carrier driven Kerr frequency comb. Higher order (>3) characteristic polynomial of intra-cavity power describing the steady-state homogeneous solution of the modified LLE are discussed in detail. The nonlinear phase detuning of the cavity has been observed through negative Kerr-tilt. We also find an analytical expression for the steady-state FCD induced detuning. Expression of MI-gain in presence of all nonlinear losses is found and the threshold detuning and the range of normalized pump detuning to initiate MI are discussed. Our theoretical study is a step towards predicting comb dynamics in the realistic cases where all the nonlinear losses and higher order dispersions are present.

**X. Acknowledgement**

Vishwatosh Mishra acknowledges financial support under CSIR SRF-scheme from CSIR, Govt. of India. Authors would also like to thank Prof. Samudra Roy, Assistant Professor, Department of Physics, Indian Institute of Technology, Kharagpur (IIT Kharagpur) and Dr. Jae K. Jang, Post-doctoral fellow, Columbia University for their valuable discussions and suggestions related to the theoretical modeling of frequency-comb generation.